\documentclass[twocolumn,showpacs,preprintnumbers,amsmath,amssymb,aps]{revtex4}

\usepackage{graphicx}
\usepackage{dcolumn}
\usepackage{bm}

\newcommand{\ket}[1]{\left|#1\right\rangle}

\newcommand{\sandwich}[3]{\langle#1|#2|#3\rangle}
\newcommand{\be}{\begin{equation}}
\newcommand{\ee}{\end{equation}}

\begin{document}

\title{Extracting Atoms on Demand with Lasers}
\author{Bernd Mohring}%
\email{bernd.mohring@physik.uni-ulm.de}
\author{Marc Bienert}
\author{Florian Haug}
\author{Giovanna Morigi}
\author{Wolfgang P. Schleich}
\affiliation{Abteilung Quantenphysik, Universit\"at Ulm, 89069 Ulm, Germany}

\author{Mark G. Raizen}
 \affiliation{
Center for Nonlinear Dynamics and Department of Physics,\\
The University of Texas at Austin, Austin, TX 78712-1081
}%

\date{\today}

\begin{abstract}
We propose a scheme that allows to coherently extract cold atoms from a
reservoir in a deterministic way. The transfer is achieved by means of
radiation pulses coupling two atomic states which are object to different trapping
conditions. A particular realization is proposed, where one state has zero magnetic moment and is confined by a dipole trap, whereas the other state with non--vanishing magnetic moment is confined by a steep microtrap potential.
We show that in this setup a predetermined number of atoms can be transferred from
a reservoir, a Bose--Einstein condensate, into the collective quantum state of the steep trap with 
high efficiency in the parameter regime of
present experiments.
\end{abstract}


\maketitle

\section{\label{sec:introduction}Introduction}
Recent experimental progress with ultracold atomic gases has opened exciting
directions in the study of many body systems, aiming at the full coherent
control of structures of increasing complexity~\cite{Bloch04}. Applications,
like realizations of quantum information protocols, are very promising and
presently investigated~\cite{hansel01,Bloch04,alber,Mandel03}. Moreover, these systems allow to study the
frontiers between the quantum and the classical world~\cite{Zurek03}.

In this perspective increasing interest has lately been attracted by the
possibility of disposing single or few cold atoms by deterministic
extraction from a reservoir, thereby allowing to control and manipulate them.
This realization of tweezers applied to quantum objects has been called
{\it quantum tweezers}. A recent proposal exploited tunnelling from a condensate to
a moving quantum dot in order to realize this scenario~\cite{diener02}.

\begin{figure}
  \begin{center}
    \includegraphics[width=8.6cm]{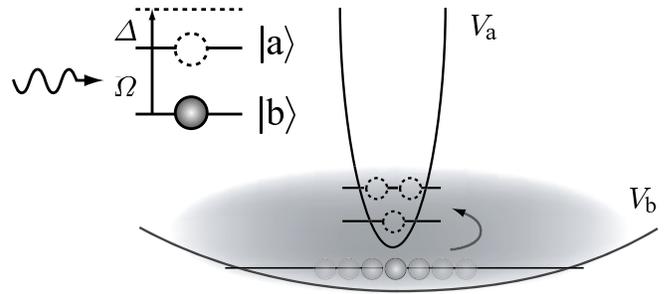}
  \end{center}
\caption{%
Extracting atoms on demand from a Bose--Einstein condensate (BEC):
Atoms are transferred from the ground state of a BEC, confined by the potential $V_b$, into the one-- or two--atom ground state of
a steep potential $V_a$ (the tweezers).
The horizontal lines indicate the corresponding energy levels (in arbitrary units).
The transfer is implemented by radiative coupling the electronic states $|b\rangle$ and $|a\rangle$, undergoing different trapping conditions.
The number of atoms transferred into the tweezers is controlled by spectrally resolving the energy splitting between the one-- and two--atom ground states of the tweezers.
}
\label{Fig:1}
\end{figure}
In this work we discuss an implementation of quantum tweezers, where trapping potentials, whose steepness depends on the atomic magnetic moment, are coupled by means of radiation. This procedure has been utilized for manipulating cold atomic clouds~\cite{Ketterle03,Mandel03b}, and it has been proposed for engineering collective states of atoms in optical lattices~\cite{Rabl02}.
Here, it is applied in order to coherently load atoms from a condensate into a steep trap by suitably coupling the atom electronic states. A sketch of the setup is shown in Fig.~\ref{Fig:1} and can be summarized as follows.
Atoms are initially in a hyperfine state of zero magnetic moment, which we denote by $\ket{b}$, where the atomic center--of--mass is confined by a dipole trap, and form a Bose--Einstein condensate (BEC).
The state $\ket{b}$ is coupled with radiation to the hyperfine state $\ket{a}$, which has non--vanishing magnetic moment and is subject to the steep potential of a magnetic trap.
The atoms are selectively transferred by means of radiative coupling between the condensate and a collective quantum state
of the tweezers. The two traps can be found in the same spatial region,
and then coherently separated spatially after the transfer has occurred~\cite{Mandel03b,Meschede03}. 

We investigate the dynamics and efficiency of quantum tweezers
using radiation and show that their efficiency for transferring a certain number of atoms into the quantum state of a trap can be larger than 99\% in experimentally accessible parameter regimes~\cite{treutlein04,Vuletic04,Schmiedmayer03}. We study three 
implementations of atom transfer into the quantum tweezers:
(i) by means of the adiabatic passage by
adiabatically ramping the radiation frequency across the
resonance~\cite{Eberly} (ii) by combining adiabatic and diabatic passages by
means of laser pulses~\cite{vitanov01}, and (iii) by resonant coupling 
using pulses of well defined area. The various techniques are compared and
their efficiency is discussed in the parameter regime of experiments with
microtraps~\cite{Reichel02}.

This article is organized as follows. In section~\ref{Sec:2} the theoretical
model is introduced and the relevant parameters are identified. In
section~\ref{sec:tweezing} the efficiency of the transfer for different realizations of the radiation coupling between the two hyperfine states is discussed. In
section~\ref{sec:conclusion} the conclusions are drawn and outlooks are
discussed. In the appendix a detailled analysis of the parameter regime, where the quantum tweezers can be implemented, is presented.

\section{\label{Sec:2}Theoretical Description}

We consider $N$ ultracold atoms of mass $M$ whose relevant internal degrees of
freedom are the stable hyperfine states $\ket a$ and $\ket b$. The atomic
center of mass experiences at position ${\bf x}=(x,y,z)$
a harmonic potential $V_j({\bf x})$
of the form
\be
V_j({\bf x})=\frac{1}{2}M\left(\nu_{jx}^2x^2+\nu_{jy}^2y^2+\nu_{jz}^2z^2\right)
\ee
where $\nu_{jx}$, $\nu_{jy}$, $\nu_{jz}$ are the frequencies along each cartesian
direction and $j=a,b$ labels the atomic state. The potential $V_a({\bf x})$ is steeper than
$V_b({\bf x})$, namely the frequencies $\nu_{ax},\nu_{ay},\nu_{az}\gg
\mbox{min}(\nu_{bx},\nu_{by},\nu_{bz})$. At time $t=0$ the atoms are all
in the state $\ket b$ and constitute a Bose--Einstein condensate.
The aim is to coherently transfer a predetermined number of atoms from the condensate,
which acts as a reservoir, into a collective quantum state of the atoms in the steep
potential $V_a({\bf x})$.

Transfer of atoms into the quantum tweezers are implemented by suitably driving the states $\ket{a}$ and $\ket b$ with radiation pulses, which can be in the microwave or optical regime, thereby implementing a Raman transition. Coherent and selective transfer is achieved provided the radiation pulses are sufficiently short, so that incoherent dynamics during the
interaction is negligible, and sufficiently long to spectrally resolve a selected atomic
transition. This latter statement will be quantified in Sec.~\ref{Sec:IIb}.
In the frame rotating at the atomic dipole resonance frequency, the relevant dynamics describing the interaction of the atom with radiation is given by the Hamiltonian $\mathcal{H}$, which we
write as \be \mathcal{H}= \mathcal{H}_a +
\mathcal{H}_b + \mathcal{H}_{\rm sc} + \mathcal{H}_{\rm int}(t)\,.
\label{eq:Hamiltonian} \ee Here, the terms $\mathcal{H}_a$ and $\mathcal{H}_b$
describe the dynamics of the atom's center--of--mass in the states $\ket a$ and
$\ket b$, respectively, and have the form
\begin{multline}
\label{H:i} \mathcal{H}_j=\int d{\bf x}\, \psi_j({\bf x})^{\dagger}
                \left[-\frac{\hbar^2}{2M} \nabla^2 + V_{j}({\bf
                    x})\right]\psi_j({\bf x})\\
              +\frac{1}{2}g_{jj}\int d{\bf x}\, \psi_j({\bf
               x})^{\dagger}\psi_j({\bf x})^{\dagger}
\psi_j({\bf x})\psi_j({\bf x})\,,
\end{multline}
where $\psi_j({\bf x})$ ($\psi_j({\bf x})^{\dagger}$) is the field operator annihilating (creating)
an atom at position ${\bf x}$ in the internal state $\ket{j}$, the term
$g_{jj}=4\pi\hbar^2a_{jj}/M$ represents the interaction strength of two--body
collisions of atoms in the state $\ket j$ and $a_{jj}$ denotes the $s$--wave
scattering length ($j=a,b$). Collisions between atoms in different hyperfine states are
described by the term \be \label{Hab}\mathcal{H}_{\rm sc}=g_{ab}\int d{\bf x}\,
\psi_a({\bf x})^{\dagger}\psi_b({\bf x})^{\dagger}
\psi_a({\bf x})\psi_b({\bf x}) \ee where $g_{ab}=4\pi\hbar^2 a_{ab}/M$ is the interaction strength,
with the $s$--wave scattering length $a_{ab}$ for collisions between one atom in state $\ket{a}$ and one atom in state $\ket{b}$. The coupling with radiation is described by the time--dependent term
\begin{multline}
\label{H:int:0}
\mathcal{H}_{\rm int}(t) =-\hbar \Delta(t) \int {\rm d}{\bf
x}\, \psi_a({\bf x})^{\dagger}
\psi_a ({\bf x})\\
+\frac 12\hbar\Omega_L(t)\int {\rm d}{\bf x}\,\left( \psi_a({\bf
x})^{\dagger}\psi_b({\bf x}) +{\rm H.c.}\right)\,,
\end{multline}
where $\Omega_L(t)$ is the real-valued Rabi frequency and $\Delta(t)$ is the
detuning of the radiation at time $t$ from the resonance frequency of the transition
$\ket{b}\to \ket{a}$.

All atoms are initially in the internal state $\ket b$ and form a condensate. We denote this collective state, the ground state of $\mathcal{H}_b$ for $N$ atoms, by $\ket 0$. By means of a radiation pulse a number $n$ of atoms (with $n\lll N$) is transferred into the lowest energy state of $\mathcal{H}_a$. We
denote by $|n\rangle$ the state of the system after the transfer, corresponding to $N-n$ atoms
in the ground state of $\mathcal{H}_b$ and $n$ atoms in the ground state of $\mathcal{H}_a$. The efficiency of the procedure is measured by the probability $P_{0\to n}(T)$, which corresponds to the probability at time $T$, at the end of the pulse, of finding $n$ atoms in the ground state of
$\mathcal{H}_a$. It is defined as
\be
P_{0\to n}(T)=|\langle
n|U(T)|0\rangle|^2
\ee 
with
\be U(T)={\cal T}\,\exp\left(\frac{1}{{\rm
i}\hbar}\int_0^T{\rm d}\tau \mathcal{H}(\tau)\right)
\ee and ${\cal T}$
indicates the time ordering. In section~\ref{sec:tweezing} we describe
strategies of varying the coefficients $\Delta(t)$ and $\Omega_L(t)$ in ${\cal
H}_{\rm int}$ as a function of time in order to achieve unit efficiency.

\subsection{Basic assumptions and approximations}

In order to study the dynamics and the efficiency of the tweezers, we decompose the field operator in a convenient basis. Here, we use the Fock decomposition of the field operator $\psi_a({\bf x})$, namely \cite{lewenstein94}
\begin{eqnarray}
\psi_a({\bf x})=\sum_{\vec{n}_a}\phi_{a}^{\vec{n}_a}({\bf x})a_{\vec{n}_a}
\end{eqnarray}
where ${\vec{n}_a}$ labels the excitations of the harmonic oscillator with Hamiltonian $p^2/2M + V_a({\bf x})$, $\phi_{a}^{\vec{n}_a}({\bf x})$ is its wave function and $a_{\vec{n}_a}$ is the field operator creating an atom in the corresponding state $\ket {\vec{n}_a}$. 
We remark that the state~$\ket {\vec{n}_a}$ describes single particle eigenstates, hence they are not the eigenstates of $\mathcal{H}_a$ when two or more (interacting) atoms are inside the tweezers. Nevertheless, in this case it still provides a plausible ansatz for estimating the order of magnitude of the physical parameters. 

Since the number of atoms which are extracted from the condensate is negligibly small compared with $N$, we replace the field operator for atoms in the hyperfine state~$\ket b$ by a scalar. At $t=0$ the atoms are assumed to be in the condensate ground state with macroscopic wave function $\phi_b({\bf x})$
and chemical potential $\mu$. These quantities are assumed to be constant and negligibly affected by the transfer of atoms into the tweezers. The condensate is hence a reservoir of atoms at given density $n_b({\bf x})=N|\phi_b({\bf x})|^2$. The condensate excitations are accounted for by the (low-energy) collective modes at frequencies $\omega_q$, and the part of Hamiltonian ${\mathcal H}_b$ which may limit the efficiency of the quantum tweezers takes the form ${\mathcal H}_b=\sum_q\hbar\omega_q b^{\dagger}_qb_q$
where $b_q$, $b^{\dagger}_q$ are the annihilation and creation operators of a phonon of energy $\hbar\omega_q$. In addition, quantum tweezers overlap spatially with the condensate, and collisions between the atom inside the tweezers and the condensate, described by Eq.~(\ref{Hab}),  can be detrimental for the coherence of the process. Their influence on the tweezers efficiency is discussed in Sec.~\ref{Sec:IIb}. Below, we treat the Hamiltonian term~(\ref{Hab}) as a small perturbation, whose effect is to introduce small shifts of the energy levels.

In the situations we consider the condensate ground state is coupled
(quasi-)resonantly with the ground state of ${\mathcal H}_a$ for $n$-atoms and 
one or two atoms at a time are transferred into the steep
potential. The states which are relevant to the dynamics are denoted by $\ket
0$, $\ket 1$ and $\ket 2$, corresponding to 0,1,2 atoms in the ground state of
the quantum tweezers, respectively. Their energy is 
$E(n)=\sandwich{n}{\mathcal H}{n}$ with $n=0,1,2$. By setting $E(0)=0$, they
take the form \be \label{eq:energysum} E(n)=E_{a}(n)+E_{\rm sc}(n)-n\mu\,, \ee
where $E_a(n)$ is the energy of $n$~atoms in the ground state of the
Hamiltonian $\mathcal{H}_a$ and $E_{\rm sc}(n)$ is the interaction energy due
to the term~(\ref{Hab}), while the term $-n\mu$ accounts for the extraction of
$n$ atoms from the condensate. The explicit expressions for $E_{a}(n)$ and $E_{\rm sc}(n)$ are
found from the overlap integral between the condensate wave function and the
wave function describing the ground state of $\mathcal{H}_a$ for $n$ atoms. We
evaluate them by approximating the latter with the wave function of $n$
non--interacting atoms in the ground state of the oscillator, and find
\begin{align}
E_{a}(n)&= n\left(\frac12\sum_{j=x,y,z}\hbar\nu_{aj}-\hbar\Delta\right) +\Delta
E_n
\label{eq:energya}\\
\intertext{and} E_{\rm sc}(n)&\approx g_{ab}nN\int {\rm d}{\bf x}
\,|\phi_a({\bf x})|^2|\phi_b({\bf x})|^2\,.
\end{align}
Here, we have denoted by $\phi_a({\bf x})$ the state $\phi_a^{\vec{n}_a}({\bf x})$ with $\vec{n}_a=(0,0,0)$
and $\Delta E_n$ is an energy shift due to particle--particle interaction in
state~$\ket{a}$, which can be decomposed into the
terms $\Delta E_n=\Delta E_{\rm
coll}+\Delta E_{\rm dipole}(t)$. The term $\Delta E_{\rm coll}$ accounts for the
collisions of atoms in the ground state, which we estimate for $n=2$ as
\be
\label{Delta E} \Delta E_{\rm coll}
\approx g_{aa}\int {\rm d}{\bf x}\, |\phi_a({\bf x})|^4\,,
\ee
while the term $\Delta E_{\rm dipole}(t)$ is due to
dipole--dipole radiative interaction between the atoms inside the tweezers and it
is proportional to the dissipative rate associated with the radiative transfer~\cite{Cirac93}.

The coupling due to radiation between the states $\ket n$ and $\ket{n+1}$ is described by the matrix elements $\sandwich{n}{\mathcal{H}}{n+1}=\sandwich{n+1}{\mathcal{H}}{n}^\ast$, with $n=0,1$. We
denote by
\be
\Omega_n^{(1)}=\frac{2}{\hbar}\sandwich{n}{\mathcal{H}}{n+1}
\label{eq:Eoffdiagonal}
\ee
the corresponding Rabi frequency, which takes the explicit form
\be
\label{H:int} \Omega_n^{(1)}=\Omega_L\sqrt{N(n+1)}\int {\rm d}{\bf
x}\,\phi^\ast_b({\bf x})\phi_a({\bf x})\,. 
\ee
We assume that atoms are
extracted from the center of the condensate, which here corresponds with the center of
the tweezers. For a very steep tweezers trap, whose size is much smaller
than the size of the condensate, the integral in
Eq.~(\ref{H:int}) is approximated by \be \Omega_n^{(1)}\approx\Omega_L\sqrt{n_b(n+1)}\int {\rm d}{\bf
x}\,\phi_a({\bf x})\,. \ee where $n_b=n_b(0)$ is the density at the
centre of the condensate.

The resonant coupling between the states $\ket 0$ and $\ket 2$ can be characterized by the Rabi frequency 
\begin{equation} 
\Omega_2^{(2)}=\frac{\Omega_1^{(1)} \Omega_2^{(1)\ast}}{E_2/2\hbar}\,, \label{eq:Eoffdiagonal:2}
\end{equation}
where the denominator corresponds to the value of the detuning for which the two states are resonantly coupled, and
\be
E_2=\sum_{j=x,y,z}\hbar\nu_{aj}+\Delta E_2 + E_{\rm sc}(2)-2\mu\,.
\ee
The coupling, Eq.~\eqref{eq:Eoffdiagonal:2}, has been evaluated in the limit $\Omega_1^{(1)}\ll E_2/\hbar$ in second order perturbation theory \cite{Cohen}.

For later convenience, we denote by $\Delta=E_1/\hbar$ the detuning for which the states $\ket 0$ and $\ket 1$ 
are resonantly coupled, with
\be
E_1=\frac12\sum_{j=x,y,z}\hbar\nu_{aj}+E_{\rm sc}(1)-\mu
\ee

\subsection{Values of the parameters}
\label{Sec:IIb}

In this section we estimate the order of magnitude of the energy terms and check
the consistency of our assumptions for experiments where the tweezers trap is a
magnetic microtrap. In particular, we consider the parameter regimes discussed
in~\cite{Reichel02}.

We first focus on the spectrum of one atom inside the tweezers. According to
Eqs.~\eqref{eq:energysum}--\eqref{eq:energya}, the energy of the vibrational
ground state~ $\ket 1$ reads $E(1)=E_a(1)+E_{\rm sc}(1)-\mu$ with
\be \label{eq:energye1} E_a(1)
= \frac{3}{2}\hbar\nu_a -\hbar\Delta\,. \ee
For $\nu_a=2\pi\times 30\, \rm kHz$ and $\nu_b=2\pi \times 100\,\rm Hz $
and a condensate of $N\approx 10^{3}$ $^{87}$Rb atoms with
density $n_b\approx 3\cdot 10^{13}$ atoms/cm$^3$, one finds $\mu/\hbar\approx 1.8\,\rm kHz$
while $E_{\rm sc}(1)/\hbar$ is orders of magnitude smaller. For the given value of $\nu_a$
the size of the tweezers ground state wave function is $a_{\rm Tweezer}\approx
60\,\rm nm$. Moreover, the energy distance between the ground and first excited
state is of the order of $\hbar\nu_a$. Thus, the frequency of the steep trap
sets a fundamental limit for selectively addressing the ground state of the
one--atom tweezers, thereby avoiding the excitation of other single-atom states. 

The ground state of two atoms inside the tweezers has energy $E(2)=E_a(2)+E_{\rm sc}(2)-2\mu$ 
with \be \label{eq:energye2} E_a(2)= 3\hbar\nu_a-2\hbar\Delta+\Delta
E_2 \,, \ee where $E_{\rm sc}(2)\sim 2E_{\rm sc}(1)$. The term $\Delta
E_2=\Delta E_{\rm coll}+\Delta E_{\rm dipole}$ describes the energy shift due
to atom--atom interaction inside the tweezer. The term due to collisions is
$\Delta E_{\rm coll}/\hbar\approx 2\pi\times 2\,\rm kHz$.
When the coupling between the hyperfine states is achieved with coherent
Raman transitions and radiation in the optical range, the term due to dipole--dipole
interaction is of the order of the excitation rate of the state~$\ket{a}$ and it is
negligible for the considered setup. Hence, the frequency of the steep trap is the largest
frequency scale, which determines also the size of the energy separation
between the ground and first excited state of the two--atom tweezers.

From Eq.~(\ref{eq:energya}) it is visible that selective addressing
of the states $\ket 1$ and $\ket 2$ requires to resolve frequencies of the order
of $\Delta E_2/\hbar$, namely the interaction energy of the atoms inside the tweezers.
This sets a fundamental limit on the parameters for implementing the transfer, and consequentely
on the time for implementing the transition. 

As in this parameter regime $\Delta E_2\gg E_{\rm sc}$, it is justified to treat the Hamiltonian term, Eq.~(\ref{Hab}), as a small perturbation. Neglecting the coupling of the state $\ket a$ with the excitations of the condensate is more delicate. This coupling can be neglected when the pulses spectrally resolve the condensate excitations. This assumption may be reasonable when the trap frequency $\nu_b> 2\pi\times 1\,\rm kHz$. In this work we will neglect this coupling, assuming a parameter regime where these excitations can be spectrally resolved. For smaller values of $\nu_b$, however, its effect must be taken into account. Its systematic study in such regime will be object of future investigations.

We discuss now the approximation on the ground state wave function of $\ket{n}$.
It is in fact questionable whether one can approximate the wave function of the
state $\ket n$, with $n\ge 2$, with the product of single particle wave
functions. We note, however, that the size of
the ground state wave function in the steep traps we consider here is still larger than the
$s$--wave scattering length. Numerical studies have shown
that two--body $s$--wave scattering is still a reasonable approximation in relatively
steep traps~\cite{tiesinga00,bolda02,bolda03}. Hence, although the ansatz we use is not reliable
for exact results, it is still plausible for gaining insight into the efficiency of the tweezers.
Nevertheless, it is understood that the application of the schemes 
discussed below to a certain experiment requires the accurate knowledge of the
relevant parameters, which have to be evaluated for the specific atomic species
and trapping conditions.

\section{\label{sec:tweezing}Tweezing Atoms with lasers}

In this section we discuss three methods of shaping radiation pulses in order
to transfer a definite number of atoms from the condensate into the microtrap.
The first method uses adiabatic ramping of the frequency of the radiation
coupling $\ket{n-1}\to \ket n$ (adiabatic passage)~\cite{Eberly,Rabl02}. The second
method is based on the combination of two pulses, which induce a sequence of
adiabatic and diabatic passages (SCRAP)~\cite{rickes00}. The
third method uses resonant pulses with well defined pulse area~\cite{Eberly}.

\subsection{\label{sec:passage}Adiabatic Passage}

We consider the transfer of $n$ atoms from the condensate into the quantum tweezers
by adiabatically ramping the frequency of radiation coupling the two atomic states. The
frequency of the pulse, namely the detuning~$\Delta(t)$ in Eq.~(\ref{H:int:0}), is slowly
varied as a function of time from the initial value $\Delta_i$ to the final
detuning $\Delta_f$, allowing the system to follow adiabatically the evolution.

\begin{figure}
  \begin{center}
    \includegraphics[width=8.6cm]{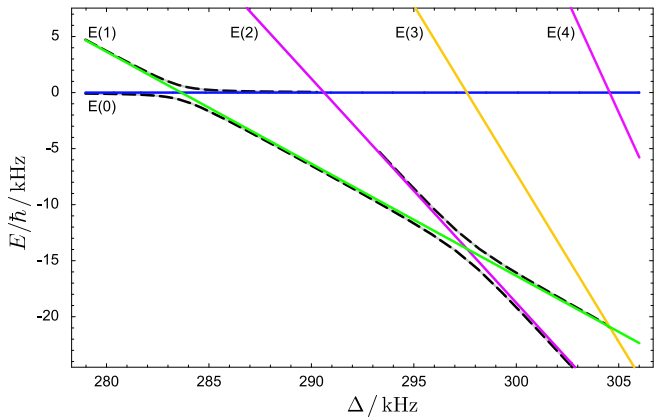}
  \end{center}
\caption{%
Extraction by adiabatic ramping the laser detuning $\Delta$ through atomic resonance.
The color lines show the energies $E(n)$, when $n$ atoms are inside the tweezers ground state, as a function of the detuning.
The energies $E(n)$ cross the energy $E(0)$, horizontal line, at different values of $\Delta$,
corresponding to the interaction energy of $n$ atoms inside the tweezers.
The dashed lines are the dressed state energies, obtained by radiative coupling between the BEC and the tweezers.
When the atoms are all initially in the BEC and the detuning is ramped adiabatically, the system dynamics follow correspondingly the lower dashed line. Transfer of one atom into the ground state of the tweezers is achieved by ramping through the avoided crossing between $E(0)$ and $E(1)$.
\label{fig:energy}}
\end{figure}

Figure~\ref{fig:energy} displays the energy eigenvalues of $\mathcal{H}$
as a function of the detuning~$\Delta$. 
Adiabatically sweeping the
detuning through these resonances corresponds to remain in an instantaneous
eigenstate of the Hamiltonian~$\mathcal{H}$. 
Transfer of one atom is implemented by ramping the detuning between the values $\Delta_i$ and $\Delta_f$ in Fig.~\ref{fig:energy}, corresponding to 
change the energy eigenstate along the dashed curve.
Adiabatic evolution is preserved when the frequency ramping rate $\dot{\Delta}$ fulfills the relation
\be
\label{Rate:Ramping}
|\dot\Delta(t_0)| \ll \frac{\pi\left(\Delta
E_2/\hbar\right)^2}{2|\ln \left (1-P_0\right )|}\,,\ee
where $\Delta E_2$ is the two-atom interaction energy and $P_0$ is the threshold transfer probability, such that for $P_{0\to 1}>P_0$ the transfer into the tweezers has been successfully implemented. The derivation of inequality~(\ref{Rate:Ramping}) is shown in Appendix A.

\begin{figure}
  \begin{center}
    \includegraphics[width=8.6cm]{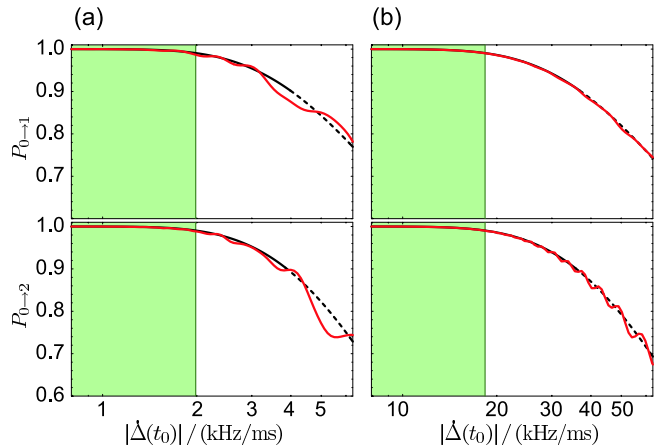}
  \end{center}
  \caption{%
Transfer efficiency $P_{0\to 1}$ (top) and $P_{0\to 2}$ (bottom) as a function of the ramping rate $|\dot\Delta|$ for (a) $\nu_a=2\pi\times 30\,\rm kHz$ and $\Omega_L=4\,\rm kHz$ (b) $\nu_a=2\pi\times 100\, \rm kHz$ and $\Omega_L=30\,\rm kHz$. The red curves show the numerical result,
the black curves display the theoretical predictions of the Landau--Zener formula. The shaded areas denote the regions, where the theoretical success probability, according to Eq.~\eqref{eq:LandauZener}, is above 99\%. Efficient transfer can be achieved in few milliseconds.
}%
\label{fig:passage}%
\end{figure}
The upper quadrants of Fig.~\ref{fig:passage} display the transfer efficiency $P_{0\to 1}$ as a
function of the rate of ramping $|\dot{\Delta}|$ for two values of the microtrap frequencies $\nu_a=2\pi\times30\,\rm kHz$ and $\nu_a=2\pi\times 100\,\rm kHz$. Here the red curve has been obtained from numerical simulations, and the shaded region denotes the range where the transfer probability is larger than 0.99.
The black curve corresponds to the prediction of the Landau--Zener formula in the adiabatic regime, namely for large transfer efficiencies~\cite{Landau,Zener,suominen92,vitanov01} and discussed in appendix A. This takes the form
\be P_{0\rightarrow 1}^{(LZ)}= 1-e^{-\alpha_{\rm ad}}
\label{eq:LandauZener} \ee with $\alpha_{\rm ad}=\pi\delta^2/2\hbar^2 | \dot\Delta(t_0)|$ the adiabaticity parameter and $\dot{\Delta}(t_0)$, $\delta$ the rate of ramping and the energy splitting, respectively, at the avoided crossing between $E(0)$ and $E(1)$. Efficiencies of the order of 99\% are achieved for transfer times of the order of the millisecond. 
The results show that faster transfer is achieved at larger trap frequencies. In fact, the interaction energy $\Delta E_2$ increases with $\nu_a$, thereby allowing for larger values of the ramping rate according to Eq.~(\ref{Rate:Ramping}).

The transfer of two atoms into the tweezers can be achieved starting from the state $\ket 0$
by adiabatically ramping the detuning $\Delta$ either (i) subsequently across the resonances $E(0)=E(1)$  and then $E(1)=E(2)$ (lower dashed line in Fig.~\ref{fig:energy}), or (ii) across the
resonance $E(0)= E(2)$.
The first case corresponds to transferring one atom at a time.
In the second case two atoms are transferred simultaneously. This latter implementation has a drawback due to the small value of the radiative coupling between the states $\ket 0$ and $\ket 2$. In fact, the corresponding Rabi frequency, estimated in Eq.~(\ref{eq:Eoffdiagonal:2}), is in general rather smaller than the Rabi frequency~$\Omega_1^{(1)}$, Eq.~(\ref{H:int}), describing the transfer of one atom into the tweezers. This smaller values leads to smaller energy splitting at the avoided crossing between the levels $\ket 0$ and $\ket 2$, therefore imposing much stricter limits on the ramping speed.
The lower quadrants of Fig.~\ref{fig:passage} depict the results of the simulation for the
sequential adiabatic transfer of two atoms into the microtrap.
The efficiencies and transfer times are comparable with the efficiencies reached for transferring a single atom, showing that two atoms can be transferred into the ground state of a tweezers trap with $\nu_a=2\pi \times 30 \,\rm kHz$ in few milliseconds. The black curve corresponds to the Landau-Zener prediction, which in the adiabatic region is to good approximation the product of the transfer probabilities at the single avoided crossings, and takes the form $P_{0\rightarrow
2}^{(LZ)}\approx P_{0\rightarrow 1}^{(LZ)}P_{1\rightarrow 2}^{(LZ)}$. Here 
$P_{1\to 2}$ is found from Eq.~(\ref{eq:LandauZener}), where the energy splitting $\delta$ is  $\hbar\left|\Omega_2^{(1)}\right|$.

\subsection{\label{sec:pulses}Adiabatic Passage using Pulses}

\begin{figure}
  \begin{center}
    \includegraphics[width=8.6cm]{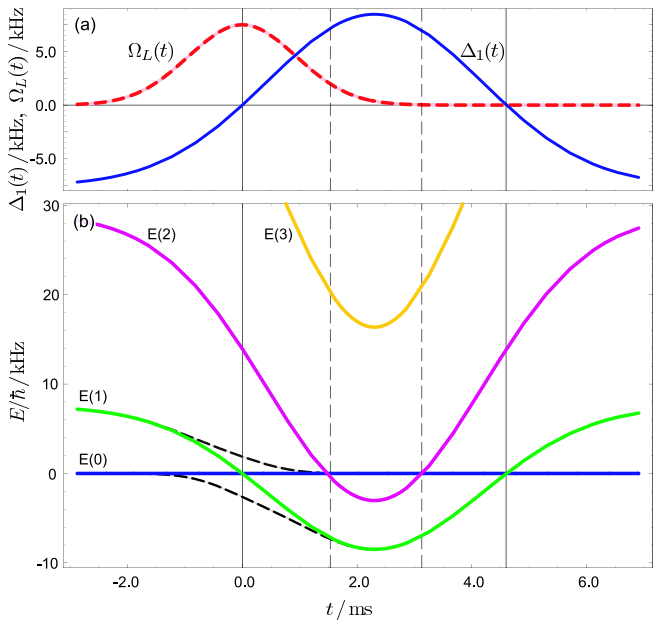}
  \end{center}
  \caption{%
Sketch of the dynamics when implementing the SCRAP technique for transferring one atom into the tweezers. 
(a) Pump pulse ($\Omega_L(t)$) and a.c.--Stark shift ($\Delta_1(t)$) as a function of time. (b) Corresponding spectrum. The color lines in (b) are the energies $E(n)$ as evaluated at the corresponding value of $\Delta_1(t)$. The dashed curves indicate the dressed state energies. The vertical solid (dashed) lines indicate the instants at which the transition $\ket 0 \to\ket 1$ ($\ket 0 \to\ket 2$) is driven resonantly. 
}%
  \label{fig:scrap}
\end{figure}

We discuss now transfer of atoms into the tweezers by means of the so-called Stark--chirped rapid adiabatic passage (SCRAP) technique~\cite{rickes00}. Here, population transfer between two quantum states is achieved by
suitably varying the Rabi frequency~$\Omega_L(t)$ and the detuning $\Delta(t)$
in Eq.~(\ref{H:int:0}), thereby combining
adiabatic and diabatic transitions. In this scheme, two pulses of finite duration
are utilized,
which are time--delayed one with respect to the other. One pulse is far--off
resonance and induces an a.c.--Stark shift $\Delta(t)$ on the transition
$\ket{b}\to\ket{a}$, while the second pulse, which we denote by `pump pulse',
couples (quasi--)resonantly the states $\ket{a}$ and $\ket{b}$ and has Rabi
frequency $\Omega_L(t)$.
The detuning is shaped in order to fulfill the resonance condition $\Delta(t)=E_1/\hbar$
at two instants of time $t_1$ and $t_2$. The shaping and
time--delay between the two pulses is such that at $t_1$ the system undergoes an
adiabatic passage of the same type as the one discussed in
Sec.~\ref{sec:passage}, while at $t_2$ the transition is diabatic, thereby
ensuring that at the end of the pulses the transfer between the states $\ket 0$
and $\ket 1$ is achieved~\footnote{In fact, if both transitions would be adiabatic
there would be no transfer at the end of the pulses as the transition at $t_2$ would reverse the
transition at $t_1$, bringing the system back to the initial state.}. 
A possible realization of the pulses as a function of time  and the corresponding dynamics
are displayed in Fig.~\ref{fig:scrap}. Here, the energy levels $E(0)$ and $E(1)$ 
cross at the instant $t_1=0$ and $t_2=\tau$. At these instants, the system
undergoes an adiabatic and diabatic transition, respectively. As a result, if initially all atoms are in the condensate, then 
finally one atom is transferred into the ground state of the tweezers.

Since at least at one avoided crossing the dynamics must be adiabatic, the bounds on the values of the Rabi frequency and on the time variation of the a.c.--Stark shift are basically the same as for the adiabatic passage in Sec.~\ref{sec:passage}. Moreover, the finite size of the pulses and the requirement that a diabatic transition takes place at one of the crossing, introduces a further parameter, namely the width of the pump pulse $T_{\Omega}$, on which the transfer efficiency depends. The parameter regime for the applicability of this technique is discussed in detail in Appendix B.

\begin{figure}
  \begin{center}
    \includegraphics[width=8.6cm]{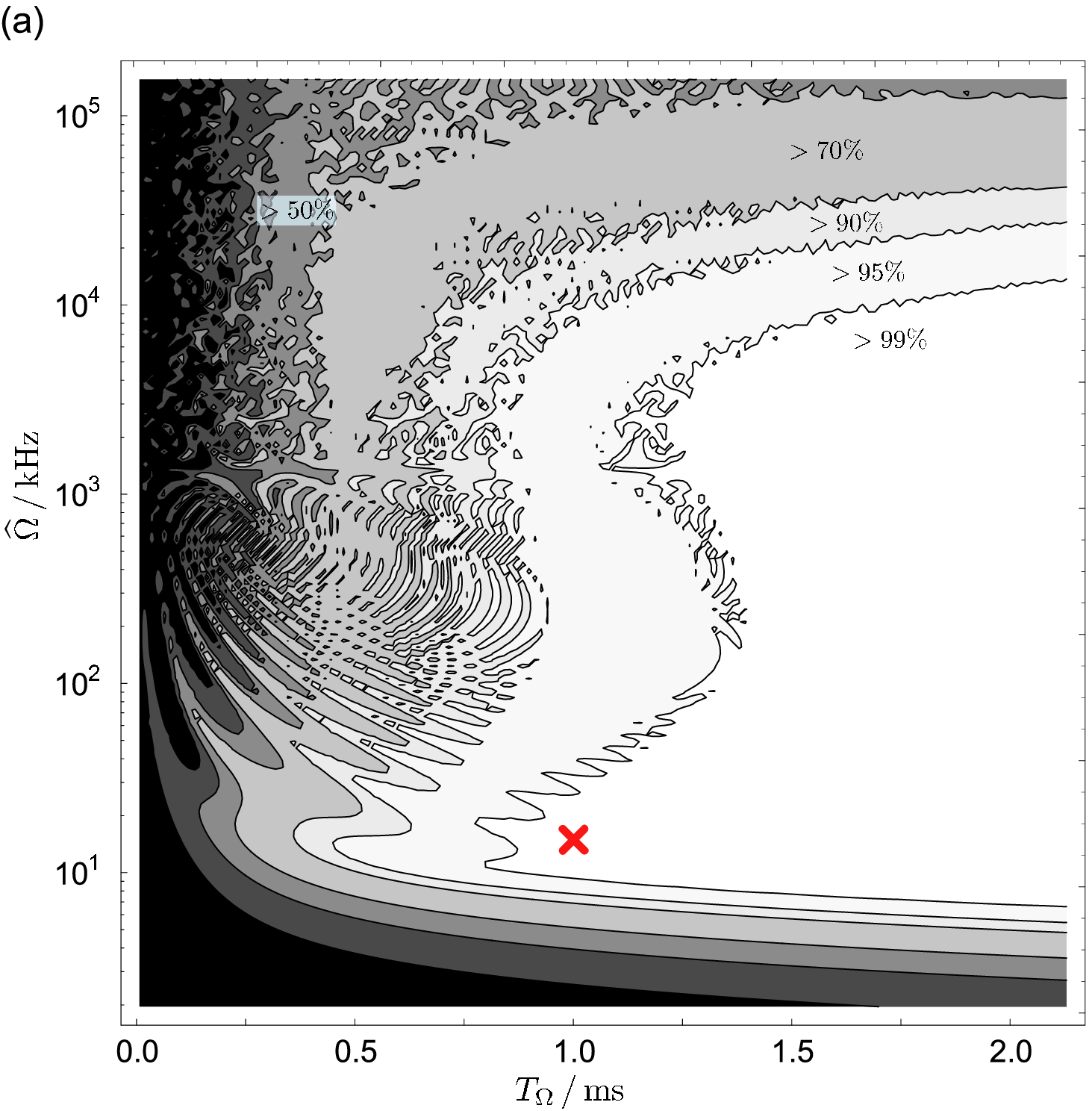}
  \end{center}
  \begin{center}
    \includegraphics[width=7.6cm]{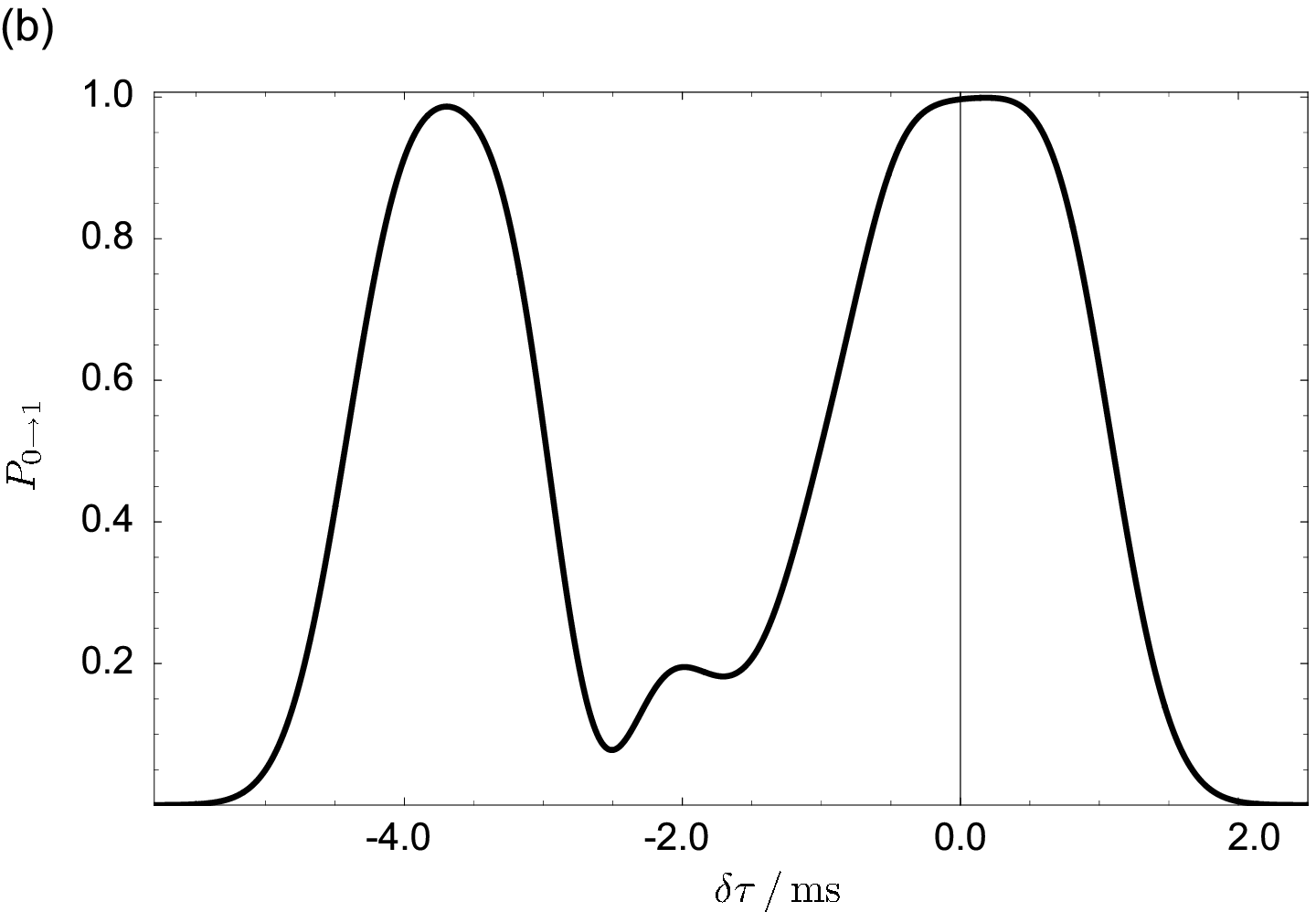}
  \end{center}
    \caption{
(a) Contour plot for the transfer probability $P_{0\to 1}$ as a function of the coupling pulse maximum $\widehat\Omega$ and width $T_\Omega$. The tweezers trap frequency is $\nu_a=2\pi\times30\,\rm kHz$ and the a.c.--Stark shift pulse has width $T_\Delta=2T_\Omega$. Pump and Detuning pulses have Gaussian envelope. The red cross indicates the parameters $\widehat\Omega=15 \rm kHz$ and $T_\Omega=1\rm ms$, which are used in evaluating the curve in (b). (b) Probability $P_{0\to1}$ as a function of the time variation~$\delta\tau$ between the instant of time at which the resonance condition is achieved by the a.c.--Stark pulse and the instant of time at which the pump pulse reaches its maximum value. The maximum at $\delta\tau\approx-4\,\rm ms$ corresponds to the dynamics for which the temporal sequence of adiabatic and diabatic passages is exchanged. The asymmetry between the two peaks is discussed in the text. 
}%
  \label{fig:scrap_result}
\end{figure}
Figure~\ref{fig:scrap_result}(a) displays the transfer efficiency as a function of $T_{\Omega}$ and $\Omega_L$, the maximum Rabi frequency of the pump pulse, as obtained from a numerical simulation for a tweezer trap with $\nu_a=2\pi\times30\,\rm kHz$. Transfer efficiencies $P_{0\to 1}>99\%$ are  achieved in the white region, corresponding to transfer times of the order of some milliseconds. High efficiencies are reached for a relatively wide range of parameters, showing that the method is robust to fluctuations of the parameters of the pulses. 
These datas have been evaluated under the condition for which the pump pulse achieves its maximum value at the same instant of time at which the detuning pulse fulfills the resonance condition, as depicted in Fig.~\ref{fig:scrap}. In this case the ideal conditions for adiabaticity are reached. The effect of a time--lag $\delta\tau$ between these two events is shown in Fig.~\ref{fig:scrap_result}(b), which reports the dependence of the transfer efficiency on $\delta\tau$. The transfer efficiency exhibits a wide plateau around $\delta\tau=0$, which is of the order of a millisecond, showing that the method is also very robust against this kind of fluctuations.
Note that the curve exhibits a second maximum at $\delta\tau\approx -4\,\rm ms$, corresponding to the case in which the pump pulse is centered at the second instant $t_2$ at which $\Delta$ fulfills the resonance condition $\Delta=E_1/\hbar$, namely to the situation in which the sequence of diabatic and adiabatic passages is reversed. Ideally, the two maxima should be symmetric. Asymmetry here arises from the fact that the dynamics at the crossing corresponding to the resonance for simultaneous transfer of two atoms into the tweezers (dashed vertical line in Fig.~\ref{fig:scrap}(b)) are not perfectly diabatic. This introduces a fundamental difference in the efficiency of transfer, which depends on whether this transition point is crossed, namely on whether the diabatic transitions occur before the adiabatic one.

\begin{figure}
  \begin{center}
    \includegraphics[width=8.6cm]{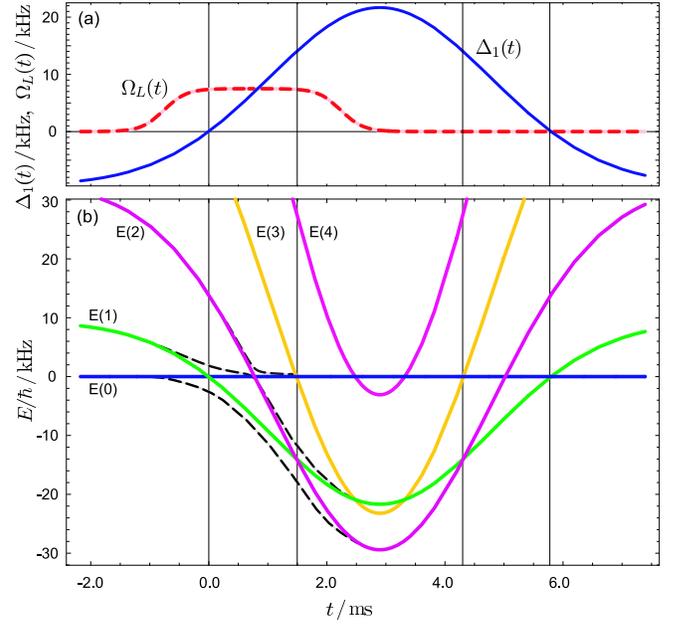}
  \end{center}
  \caption{
Sketch of the dynamics when implementing the SCRAP technique for transferring sequentially two atoms into the tweezers. (a) Pump pulse ($\Omega_L(t)$) and a.c.--Stark shift ($\Delta_1(t)$) as a function of time. (b) Corresponding spectrum. The color lines in (b) are the energies $E(n,t)$ as evaluated at the corresponding value of $\Delta_1(t)$. The dashed curves indicate the dressed state energies. 
The vertical lines indicate the instants at which the transition $\ket 0 \to\ket 1$  and $\ket 1 \to\ket 2$ are driven resonantly.
Transfer of two atoms is achieved when the dynamics start at the blue solid curve. 
}
  \label{fig:scrap:2}
\end{figure}

\begin{figure}
  \begin{center}
    \includegraphics[width=8.6cm]{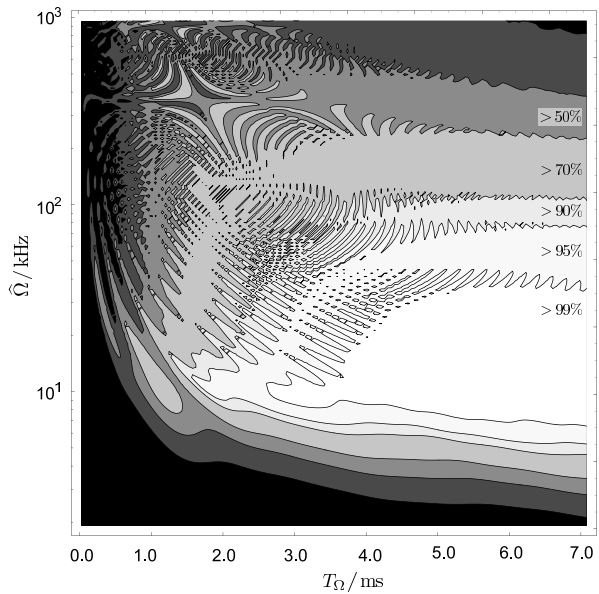}
  \end{center}
\caption{
Contour plot for the transfer probability $P_{0\to 2}$ as a function of the coupling pulse maximum $\widehat\Omega$ and width $T_\Omega$.
The tweezers trap frequency is $\nu_a=2\pi\times30\,\rm kHz$ and the a.c.--Stark shift pulse has width $T_\Delta=2T_\Omega$.
The detuning pulse has Gaussian envelope, the pump pulse is given by $\frac{\Omega_0}{2}\left(\tanh (\frac{t + 2t_{\rm ramp}}{t_{\rm ramp}})-\tanh 
 (\frac{t - T_\Omega-2t_{\rm ramp}}{t_{\rm ramp}})   \right)$, where $t_{\rm ramp}$ is the time in which the pulse is ramped up to the maximum value.
}%
  \label{fig:scrap:2result}
\end{figure}

Similarly to the adiabatic passage in Sec.~\ref{sec:passage}, transfer of two atoms by means of SCRAP is better implemented sequentially, i.e., by transferring one atom at a time. Figure~\ref{fig:scrap:2} shows a pulse sequence and the corresponding energy spectrum as a function of time for transferring sequentially two atoms inside the tweezers. The ideal dynamics follow the lower dashed line in Fig.~\ref{fig:scrap:2}(b).
The transfer efficiency as a function of $T_{\Omega}$ and $\widehat{\Omega}$ is displayed in Fig.~\ref{fig:scrap:2result}. Here, efficient transfer of two atoms, larger than $99\%$, is achieved in times of the order of several milliseconds.
The relatively broad range of values for which $P_{0\to 2}$ exceeds $99\%$ shows that the method is robust against parameter fluctuations. 

\subsection{\label{sec:discussion}Transfer by population inversion}

\begin{figure}
  \begin{center}%
    \includegraphics[width=8.6cm]{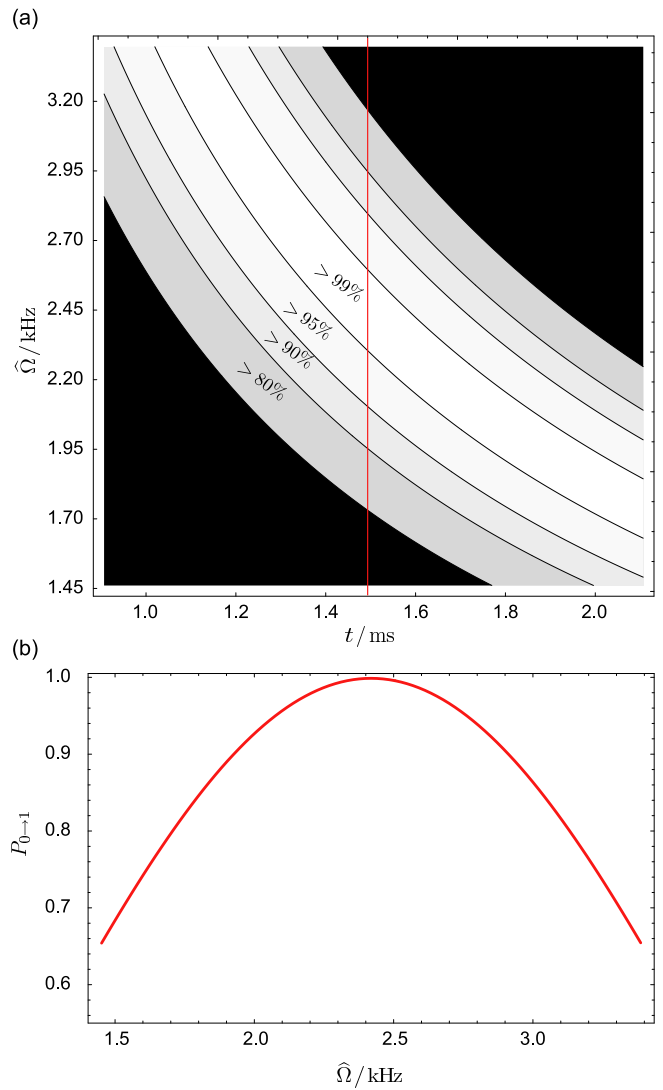}%
  \end{center}%
  \caption{%
Transfer of one atom by resonant $\pi$ rotation.
(a)~Contour plot of $P_{0\to 1}$ by means of a Gaussian pulses with $\Delta=E_1/\hbar$
as a function of the maximum pulse value $\widehat\Omega$ and of the pulse width
$T_\Omega$ for $\nu_a=2\pi\times30\,\rm kHz$. The black regions correspond to the parameter regime, where the efficiency is below 80\%.
(b)~$P_{0\to 1}$ as a function of the pulse strength $\widehat\Omega$ for constant pulse width $T_\Omega=1.5\,\rm ms$. The corresponding parameters are indicated by the vertical line in (a).
}%
  \label{fig:raman}%
\end{figure}

We finally discuss transfer of atoms into the tweezers by radiation pulses of
chosen areas, coupling resonantly the states $\ket 0$ and $\ket 1$. Perfect
transfer is achieved when~$\Delta=E_1/\hbar$ and when the 
the pulse area fulfills the relation $\int {\rm d}\tau\, \Omega_1^{(1)}(\tau)=\pi\,$.
The limitations are on the choice of the Rabi frequency, whose value is bound by the
interaction energy in order to have negligible coupling to off--resonant
states (see Appendix A), and on the pulse duration, which must be sufficiently long to guarantee
the spectral resolution of the tweezers energy levels. In Fig.~\ref{fig:raman} the
transfer efficiency $P_{0\to 1}$ is plotted for Gaussian pulses as a function
of the pulse maximum intensity and duration. Efficiencies close to unity are
realized with pulses of millisecond duration.

Efficient transfer of two atoms is achieved by sequentially sending two pulses,
each transferring one atom into the tweezers. Here, the first pulse is resonant
with the transition $\ket 0\to \ket 1$, thereby implementing a $\pi$-rotation.
A $\pi$-rotation on the transition $\ket 1\to \ket 2$ is achieved by coupling
it with a second pulse with detuning $\Delta=(E_2-E_1)/\hbar$ and suitable pulse 
area. The spectral resolution, at the level of $\Delta E_2/\hbar$, avoids that during  
the second pulse the atom in the tweezers is transferred back to the 
condensate, as the transition $\ket{0}\to\ket{1}$ is far--off resonance.
Transfer efficiency exceeding $99\%$ are achieved on time scales of the order
of several milliseconds.

\section{\label{sec:conclusion}Discussion and Conclusions}

We have investigated the feasibility of a scheme that employs radiation, thereby achieving the deterministic extraction 
of atoms from a Bose--Einstein condensate and their transfer into the quantum state of a steep trap.
We have shown that a high transfer efficiency can be achieved in times of the order of some milliseconds in the parameter regime of present experiments with microtraps~\cite{Reichel02}. The radiation used for implementing the extraction can be either in the microwave regime, thus coupling directly, say, a magnetic dipole transition like
in~\cite{Ketterle03,Mandel03b}, or in the optical domain, thereby implementing Raman
transitions. Their application depends on the specific details of the system. 

The results here presented refer to ideal conditions, where at $t=0$
all atoms are in the condensate ground state and the coupling with the
other excited states of the condensate can be neglected. The presence of non--condensed atoms
and the coupling of the state $\ket a$ to the condensate excitations 
have not been considered. The presence of non--condensed atoms introduces a source of noise into the process, giving an indetermination in the transfer efficiency of the order of the non--condensed fraction. The coupling of the state $\ket a$ to the condensate excitations constitutes a
further source of noise which limits the efficiency of the tweezer. Its effect can be
eliminated by resolving spectrally the condensate excitations. This is possible when the
condensate is confined in a sufficiently steep trap. In this case, the condensate excitations must be taken into account when setting the parameters, as they determine the spectral resolution to be achieved and eventually the transfer duration.

Our analysis has been restricted to the transfer of one and two atoms into the
corresponding tweezers ground state. Similar
considerations apply for the transfer into one-- and two--atom excited states. In
this case, the parameters must be carefully set, so to spectrally resolve the
desired level. The process, in this case, may take advantage of symmetries,
ruling out the coupling to states which may be close but orthogonal to the
initial state. The procedure discussed in this paper can be as well extended
to the transfer of three or more atoms into the tweezers. In this case, we expect that
the scheme is efficient when transferring sequentially one atom at a time. 
When confining three or more atoms into the steep trap, however,
one must consider many--body effects which can give rise to instabilities. The 
dynamics and time--scales on which they manifest themselves depend on the trap parameters and atomic species and has to be
analysed case by case.

To conclude, the possibility of extracting atoms on demand from a condensate
realizes the deterministic coupling of an atom trap to a reservoir, thereby
finding several analogies in quantum optical systems, and opens interesting
perspectives in the study and the control of coherent matter waves.

\begin{acknowledgments}
The authors gratefully acknowledge discussions with Andrew J. Daley, Murray Holland, Peter Rabl, Jakob
Reichel, Bruce W. Shore, Philipp Treutlein and Reinhold Walser. Mark G. Raizen and Wolfgang P. Schleich thank the Humboldt Foundation for the support of a Max Planck Research Award.  
Support from the EU--network CONQUEST, the Landes\-stiftung Baden--W\"urttemberg in the framework of the Quanten\-auto\-bahn~A8,
and the National Science Foundation;  Quantum Optics Initiative, US Navy -- Office of Naval Research Grant No. N00014-04-1-0336
are acknowledged. 
\end{acknowledgments}

\begin{appendix}

\section{Limits for the application of the adiabatic passage}

In this appendix we discuss the parameter regimes for which the dynamics described in Sec.~\ref{sec:passage} apply. We first summarize the dynamics of the adiabatic passage, by discussing the transfer of one atom into the tweezers. The considerations made here can be directly extended to the transfer of $n$ atoms. We assume that the detuning is swept between the initial value $\Delta_i$ and the final value $\Delta_f$ in the interval of time $T$, such that $\Delta(0)=\Delta_i$ and $\Delta(T)=\Delta_f$. Efficient transfer from $\ket 0$ to
$\ket{1}$ is achieved when at these values of the detuning the states $\ket 0$, $\ket 1$ are eigenstates of $\mathcal{H}$ to good approximation.
The value of the Rabi frequency is bound from the constraint that the relevant dynamics must involve only the states $\ket 0$ and $\ket 1$. When these conditions are fulfilled, the dynamics can be restricted to the subspace of states $\{\ket{0},\ket{1}\}$ and the eigenvalues of the reduced Hamiltonian at a given instant $t$, $0\le t\le T$, are
\be
\epsilon_{\pm}(1,t)=\frac12\hbar\Delta_1(t)\pm\frac12\hbar\sqrt{\Delta_1(t)^2+
\left|\Omega_1^{(1)}\right|^2} \label{eq:energy_2level} \ee with
\be\label{eq:delta1} \Delta_1(t)=\Delta(t)-E_1/\hbar.\ee The corresponding
eigenstates take the form
\begin{subequations}
\begin{align}
\ket{\Phi_{+}(t)}&=\sin{\Theta(t)}\ket{0}+\cos{\Theta(t)}\ket{1}\\
\ket{\Phi_{-}(t)}&=\cos{\Theta(t)}\ket{0}-\sin{\Theta(t)}\ket{1}
\end{align}
\label{eq:states_2level}
\end{subequations}
where the mixing angle $\Theta$ is defined by the relation \be
\label{Theta}\tan
2\Theta(t)=\frac{\Omega_1^{(1)}}{\Delta_1(t)}\,,\;0\le\Theta(t)\le\pi\,. \ee

We denote by $t_0$ the instant of time at which the transition is resonantly driven,
namely when the detuning fulfills the relation $\Delta(t_0)=E_1/\hbar$. The minimum value $\delta$ of the energy splitting is reached at $t=t_0$ and takes the form
\be
\delta=\left|\epsilon_{+}(1,t_0)-\epsilon_{-}(1,t_0)\right|=\hbar\left|\Omega_1^{(1)}\right|
\label{eq:energysplitting} \ee
We now identify the parameter regime for which one may restrict the dynamics to the levels $\ket 0$ and $\ket 1$. This approximation is justified when (i) coupling to
single--particle excitations of the one--atom tweezer is negligible, i.e.\
$\left|\Omega_1^{(1)}\right|\ll\nu_a$ and (ii) when the state $\ket 2$ contributes
negligibly to the dynamics, namely when the condition
\be\label{eq:condition}\left|\Omega_1^{(1)}\right|\ll\Delta E_2/\hbar\ee
is fulfilled.
This condition on $\Omega_1^{(1)}$ ensures also a complete adiabatic transfer by ramping the detuning
across the resonance $E(0)=E(1)$ while keeping $\Omega_L$ constant. In fact, in this limit the
final detuning $\Delta_f$ can be chosen
sufficiently far away from the resonance $E_2/2\hbar$, and at the same time be sufficiently large so that at $\Delta_f$
the unperturbed states $\ket 0$ and $\ket 1$ are to good approximation eigenstates of $\mathcal{H}$.

We now evaluate the probability $P_{0\rightarrow 1}$ of transferring one atom
into the ground state of the tweezer when the dynamics can be restricted to the
levels $\ket 0$ and $\ket 1$. In the adiabatic regime the probability
$P_{0\rightarrow 1}$ can be evaluated by using the Landau--Zener formula, which at the
asymptotics, i.e., for large transfer efficiencies, takes the form
\begin{equation}
\label{eq:LandauZener2}
P_{0\to 1}^{(LZ)}=1-{\rm e}^{-\alpha_{\rm ad}}
\end{equation}
Here $\alpha_{\rm ad}$ is the adiabaticity parameter, defined as \be
\alpha_{\rm ad}=\frac{\pi\delta^2}{2\hbar^2 | \dot\Delta(t_0)|}\,,
\label{eq:adiabaticity}
\ee
The parameter $\alpha_{\rm ad}$ is proportional to the energy splitting~$\delta$,
defined in~Eq.~\eqref{eq:energysplitting}, and is inversely proportional to the
time derivative of the detuning $\Delta$ at the level--crossing. Hence, the
probability $P_{0\to 1}^{(LZ)}$ approaches unity for large values of $\alpha_{\rm ad}$,
i.e.\ for large values of the energy splitting and/or small values of the rate
$\dot{\Delta}$.

We fix now the threshold value $P_0$, such that for $P_{0\to 1}\ge P_0$ the
transfer is considered successful, and we evaluate the rate $\dot{\Delta}$, and
thus the minimum time, required for transferring successfully one atom into the
tweezers. From Eq.~(\ref{eq:LandauZener}) we find that $P_{0\to 1}^{(LZ)}\ge
P_0$ when \be |\dot\Delta(t_0)| \leq \frac{\pi\left(\delta/\hbar\right
)^2}{2|\ln \left (1-P_0\right )|}\,. \label{eq:deltadot} \ee Using
Eqs.~\eqref{eq:energysplitting} and \eqref{eq:condition} we rewrite this
condition as \be |\dot\Delta(t_0)| \ll \frac{\pi\left(\Delta
E_2/\hbar\right)^2}{2|\ln \left (1-P_0\right )|}\,. \ee Hence, the time
$\tau_1$ needed for transferring one atom into the tweezers must fulfill the
relation \be \tau_{1}=\left|\frac{\Delta_f-\Delta_i}{\dot\Delta(t_0) }\right |
\gg\frac 2\pi \left |\frac{\ln \left (1-P_0\right )}{\Delta E_2/\hbar}\right
|\,. \ee where we have taken $\left|\Delta_f-\Delta_i\right|\sim\Delta
E_2/\hbar$. With the parameters of section~\ref{Sec:IIb}, for $\Delta
E_2/\hbar=2\pi\times 2\,\rm kHz$ and a success probability $P_0=99$\%, the
transfer time must fulfill the relation $\tau_1\gg 0.2\,\rm ms$.

\section{Limits for the application of SCRAP}

In this appendix we discuss the parameter regimes for which the dynamics
described in Sec.~\ref{sec:pulses} applies. For simplicity we consider the case
in which both the pump as well as the far--off resonance pulse have Gaussian
envelope, and are given by
\begin{subequations}
  \begin{align}
     \Omega_L(t)&=\widehat \Omega \exp\left[-\frac{t^2}{T^2_\Omega}\right]\\
\intertext{and}
     \Delta_1(t)&=\Delta_{\rm offset}+\widehat\Delta \exp\left[
     -\frac{\left(t-\tau\right)^2}{T^2_\Delta}  \right]\,,\label{D:pulse}
  \end{align}
\end{subequations}
where $T_\Omega$ and $T_\Delta$ are the pulse widths, the terms $\widehat
\Omega$ and $\widehat \Delta$ denote the pulse maxima, and the offset value
$\Delta_{\rm offset}$ is such that at the instants $t_1$ and $t_2$ the
resonance conditions $\Delta(t_1)=\Delta(t_2)=E_1/\hbar$ are fulfilled.
Here, we have chosen $t_1<t_2$, so that the adiabatic passage occurs at the
first crossing, and set $t_1=0$ and $t_2=\tau$.

The system dynamics can be reduced to the two levels $\ket 0$ and $\ket 1$
provided that the relation in Eq.~(\ref{eq:condition}) is fulfilled, which set an
upper bound to the maximum value $\widehat{\Omega}$ that the pump pulse reaches.
The dynamics are adiabatic around the point $t_1$ provided that the pulses time
variation is sufficiently smooth such that the adiabaticity parameter in Eq.~(\ref{eq:adiabaticity}) is very large. 
For this reason, ideally the pump pulse should reach its maximum at $t_1$, such that the level splitting is maximum at the avoided crossing. The adiabaticity of
the transfer at $t_1$ can be evaluated by checking for which parameters the
adiabaticity parameter $\alpha_{\rm ad}$ of Eq.\ (\ref{eq:adiabaticity}) is much larger
than unity. From Eqs.~\eqref{eq:condition} and~\eqref{eq:adiabaticity} one
finds a condition for the temporal variation of the detuning. For Gauss
pulses as in Eq.~\eqref{D:pulse} this condition takes the form \be \label{SCRAP:ad}
\frac{\pi}4\frac{\Delta E_2^2}{\hbar^2}\gg\frac{\widehat\Delta\tau}{T_\Delta^2}
\exp\left [-\frac{\tau^2}{T_\Delta^2}\right ]\, \ee where here $t_0=t_1=0$.

Ideally, the adiabatic transition between two eigenstates of
$\mathcal{H}_a+\mathcal{H}_b$, say $\ket 0$ to the state $\ket 1$, is
implemented over an infinite time. As the pump pulse has a finite duration
$T_{\Omega}$, then $T_{\Omega}$ must exceed a minimum value $t_{\rm jump}$ in
order to achieve sufficiently large transition probabilities at the instant
$t_1$. Using the relation $t_{\rm jump}\sim 2\widehat{\Omega}_1/
|\dot\Delta(t_1)|$~\cite{vitanov99}, where $\widehat{\Omega}_1$ is the largest
value of the coupling between $\ket 0$ and $\ket 1$, this condition corresponds
to
\be
T_{\Omega}\ge \widehat\Omega_1^{(1)}
\frac{T^2_\Delta}{\tau\widehat\Delta}
\exp\left[\frac{\tau^2}{T^2_\Delta}\right]\,.
\ee
which fixes a
relation between the widths and maxima of the two pulses and their relative
delay. Simple algebra shows that this condition is consistent with the
condition in Eq.~(\ref{SCRAP:ad}) provided that $\Delta E_2T_{\Omega}/\hbar\gg
1$.

Finally, the transition is diabatic at $t_2=\tau$ when the pump pulse is very
small at this instant, i.e., $T_{\Delta}>T_{\rm \Omega}$, and more specifically
$\tau>T_{\rm \Omega}$. A condition for diabaticity is derivable
imposing that the adiabaticity parameter at $t_2$ is $\alpha_{\rm ad}(t_2)\ll1$, thereby
finding a bound on the time--lag $\tau=t_2-t_1$ which also depends on the form of the pulse envelopes.

\end{appendix}
\bibliography{long221204}

\end{document}